\documentclass[11pt]{article}
\usepackage{amssymb}
\usepackage{amsmath}
\usepackage{amscd}
\usepackage{latexsym}

\catcode`@=11
\renewcommand{\section}{\@startsection{section}{1}{0pt}{\medskipamount}
{\medskipamount}{\large\bf}}
\numberwithin{equation}{section}
\catcode`@=12

\def\b{\beta}
\def\g{\gamma}

\def\th{\theta}

\def\la{\lambda}
\def\m{\mu}
\def\n{\nu}
\def\r{\rho}
\def\s{\sigma}

\def\vp{\varphi}

\def\1{\bar 1}
\def\2{\bar 2}
\def\3{\bar 3}

\newcommand{\yb}{\bar{y}}

\newcommand{\zb}{\bar{z}}
\newcommand{\wb}{\bar{w}}

\newcommand{\C}{\mathbb C}
\newcommand{\R}{\mathbb R}
\newcommand{\Z}{\mathbb Z}
\newcommand{\Hbb}{{\mathbb H}}
\newcommand{\Zcal}{{\cal Z}}
\newcommand{\Acal}{{\cal A}}
\newcommand{\Ncal}{{\cal N}}
\newcommand{\Lcal}{{\cal L}}
\newcommand{\Fcal}{{\cal F}}
\newcommand{\Ecal}{{\cal E}}
\newcommand{\J}{{\cal J}}

\def\im{\mbox{i}}
\def\N2{$N{=}2$}
\def\pa{\mbox{$\partial$}}
\def\diff{\mbox{d}}
\def\tr{{\rm tr}}
\def\sfrac#1#2{{\textstyle\frac{#1}{#2}}}
\def\>{\rangle}
\def\<{\langle}
\def\+{\dagger}
\def\={\ =\ }

\begin{document}

\begin{titlepage}
\setcounter{page}{0}
\begin{flushright}
.
\end{flushright}

\vskip 2.5cm

\begin{center}

{\Large\bf
 Integrability of Vortex Equations on Riemann Surfaces
   }

\vspace{15mm}
{\Large Alexander D. Popov}
\\[5mm]
\noindent {\em Bogoliubov Laboratory of Theoretical Physics, JINR\\
141980 Dubna, Moscow Region, Russia}
\\
{Email: {\tt popov@theor.jinr.ru}}
\vspace{15mm}

\begin{abstract}
\noindent
The Abelian Higgs model on a compact Riemann surface $\Sigma$ of genus $g$ is considered.
We show that for $g>1$ the Bogomolny equations for multi-vortices at critical coupling can be
obtained as compatibility conditions of two linear equations (Lax pair) which are written down
explicitly. These vortices correspond precisely to SO(3)-symmetric Yang-Mills instantons on the
(conformal) gravitational instanton $\Sigma\times S^2$ with a scalar-flat K\"ahler metric. Thus,
the standard methods of constructing solutions and studying their properties by using Lax pairs
(twistor approach, dressing method etc.) can be applied to the vortex equations on $\Sigma$.
In the twistor description, solutions of the integrable vortex equations correspond
to rank-2 holomorphic vector bundles over the complex 3-dimensional twistor space of
$\Sigma\times S^2$. We show that in the general (nonintegrable) case there is a
bijection between the moduli spaces of solutions to vortex equations on $\Sigma$ and
of pseudo-holomorphic bundles over the almost complex twistor space.
\end{abstract}
\end{center}
\end{titlepage}

\section{Introduction}

\noindent The Abelian Yang-Mills-Higgs model on $\R\times\R^2$ at critical value of the
coupling constant (the Bogomolny regime) admits static finite energy vortex
solutions~\cite{Abrikosov:1956sx,  Nielsen:1973cs} whose existence on $\R^2$ was
proved by Taubes~\cite{Taubes:1979tm, Taubes:1979ps}. They describe magnetic flux tubes
(Abrikosov strings~\cite{Abrikosov:1956sx,  Nielsen:1973cs}) penetrating a two-dimensional
superconductor. Stability of vortices is ensured by topology~\cite{JT}.
The Taubes results for critically coupled vortices were generalized to compact Riemann
surfaces $\Sigma$~\cite{Bradlow:1990ir, GP1}. In these investigations the main attention was
devoted to study of K\"ahler geometry of the moduli space of vortices on $\Sigma$ and to dynamics
of vortices in Manton's adiabatic approximation (see e.g.~\cite{Manton:1993tt}-\cite{MS}).

Recall that besides proving the existence theorem, Taubes has also shown that the standard
vortex equations on $\R^2$ are equivalent to SO(3)-symmetric SU(2) self-dual Yang-Mills
(SDYM) equations on $\R^2\times S^2$~\cite{Taubes:1979ps}. In other words, there is a one-to-one
correspondence between SO(3)-equivariant\footnote{This means a generalized SO(3)-invariance, i.e.
invariance up to gauge transformations~\cite{Forgacs:1979zs}. Identifying $S^2=\ $SO(3)/SO(2) with
$\C P^1=\ $SU(2)/U(1), we will also speak about SU(2)-equivariance.} instantons on $\R^2\times S^2$
for the gauge group SU(2) and vortices in the Abelian Higgs model\footnote{See also the previous
work of Witten~\cite{Witten:1976ck}, who reduced the SDYM equations on $\R^4$ to vortex equations on
the hyperbolic plane $\Hbb^2$.} on $\R^2$. This correspondence works also for vortices on any compact
Riemann surface $\Sigma$ of genus $g$: there is an equivalence between vortices $(A,\phi )$
on $\Sigma$ and SO(3)-equivariant Yang-Mills instantons $\Acal$ on $\Sigma\times S^2$,
with the vortex number $N$ (first Chern number of the U(1) connection $A$ on the Hermitian
line bundle $E$ over $\Sigma$) equals the instanton number (the minus second Chern number of the
symmetric connection $\Acal$ on the rank-2 vector bundle $\Ecal$ over
$\Sigma\times S^2$)~\cite{GarciaPrada:1993qv}. We describe this correspondence in explicit
form following~\cite{Popov:2005ik}.
In fact, we consider the pure SU(2) Yang-Mills action on $\Sigma\times S^2$ and show how it
reduces to the action of the Abelian Higgs model on $\Sigma$. Then from the standard Bogomolny
argument it follows that for a fixed vortex number $N\ge 0$ the minimum of the action functional
is governed by solutions of the first-order vortex equations on $\Sigma$.

The above correspondence can be advanced further. Namely, recall that to each oriented Riemannian
4-dimensional manifold $M$ one can associate a real 6-dimensional manifold $\Zcal$, the
{\it twistor space} of $M$, which has a canonical almost complex structure $\J$~\cite{Atiyah:1978wi}.
For zero scalar curvature $R_M=R_{\Sigma}+R_{S^2}$ of $M=\Sigma\times S^2$, i.e. for\footnote{Such
manifolds have self-dual Weyl tensor~\cite{B} and give particular examples of conformal
gravitational instantons discussed e.g. in~\cite{Atiyah:1978wi, Penrose:1976js, Gibbons1}.
These instantons minimize the action quadratic in the Weyl tensor for the
metric on $M$.}
\begin{equation}\label{1.1}
R_{\Sigma}=-R_{S^2}=-\frac{2}{R^2}\ ,
\end{equation}
where $R$ is the radius of $S^2$, the almost complex structure $\J$ is integrable and $\Zcal$
becomes a complex 3-dimensional manifold~\cite{B}. Then we can pull back the instanton bundle
$\Ecal$ over $M$ to a holomorphic bundle $\hat\Ecal$ over $\Zcal$~\cite{Atiyah:1978wi, Ward:1977ta}.
We show that in this (integrable) case the vortex equations on $\Sigma$ can be obtained as the
compatibility conditions of linear differential equations defining holomorphic sections of the
bundle $\hat\Ecal\to\Zcal$.
Thus, we extend the correspondence between vortices on $\Sigma$ and Yang-Mills instantons on
$\Sigma\times S^2$ further and show that for $g>1$ the vortex equations\footnote{The condition
$g>1$ is needed here since (\ref{1.1}) can be satisfied only for Riemann surfaces of genus
$g>1$.} on $\Sigma$ are equivalent to the equations defining an integrable holomorphic structure
on the smooth complex vector bundle $\hat\Ecal$ over the complex 3-dimensional twistor space
$\Zcal$.

The picture is different for the cases $\Sigma =S^2\ (g=0)$\footnote{Note that the manifold
$M=S^2{\times}S^2$ with equal scalar curvatures of two-spheres is an Einstein manifold which is
considered as a gravitational instanton in Euclidean quantum gravity, see e.g.~\cite{Gibbons2,
Gibbons3}.}, $\Sigma =T^2\ (g=1)$, $\Sigma$ with $g>1$ and $R_{\Sigma}\ne -R_{S^2}$ as well as
for the noncompact case of $\R^2$. In all these cases an almost complex structure on the twistor
space $\Zcal$ of $M=\Sigma\times S^2$ is {\it not integrable} as well as an almost complex
structure on the vector bundle $\hat\Ecal\to\Zcal$ which is the pull-back of an instanton
bundle $\Ecal\to M$. We argue that $\hat\Ecal$ with the pulled-back connection
$\hat\Acal =\pi^*\Acal$, for $\pi : \Zcal\to M$, has the structure of Bryant's
{\it pseudo-holomorphic bundle}~\cite{Bryant} with the curvature $\hat\Fcal$ of type (1,1).
Furthermore, we show that the vortex equations on $\Sigma$ (including the case of $\R^2$)
are equivalent to the SU(2)-equivariant {\it pseudo-Hermitian-Yang-Mills equations} on the almost
complex twistor space $\Zcal$. However, the vortex equations in all these cases cannot be
written as compatibility conditions of linear (Lax) equations. In the particular case of $g>1$
and equality (\ref{1.1}), all almost complex structures become integrable, Lax pair appears and
the prefix ``pseudo-" disappears.

\vspace{5mm}

\section{Scalar-flat gravitational instantons and twistors}

Here we consider K\"ahler metrics on a product $M=\Sigma_{1}\times \Sigma_{2}$
of two Riemann surfaces of genera $g_1$ and $g_2$ and the twistor description
of such four-dimensional Riemannian manifolds.

\noindent
{\bf Riemann surfaces.} Consider a compact Riemann surface $\Sigma$ of genus $g\ge 0$
 with the metric and the volume form given in local (conformal) coordinates $w, \wb$ by
\begin{equation}\label{2.1}
\diff s^2 \equiv \diff s^2_{\Sigma}= 2\,g_{w\wb}\,\diff w\,\diff \wb
\end{equation}
and
\begin{equation}\label{2.2}
\omega\equiv \omega_{\Sigma}^{} = \im\,g_{w\wb}\,\diff w\wedge\diff \wb\ .
\end{equation}
Since $\Sigma$ is a K\"ahler manifold, it follows that
\begin{equation}\label{2.3}
\Gamma^w_{ww} = 2\,\pa_w\log\rho\quad\mbox{and}\quad
\Gamma^{\wb}_{\wb\wb} = 2\,\pa_{\wb}\log\rho\quad\mbox{with}\quad
\rho^2:=g_{w\wb}\ ,
\end{equation}
\begin{equation}\label{2.4}
R_{w\wb} = -2\pa_w\pa_{\wb}\log\rho =\varkappa\, g_{w\wb}\ ,
\end{equation}
where $\Gamma^w_{ww}$ and $\Gamma^{\wb}_{\wb\wb}$ are nonvanishing components
of the Christoffel symbols and $R_{w\wb}$ is a component of the Ricci tensor.

For the scalar curvature of $\Sigma$ we have
\begin{equation}\label{2.5}
R^{}_{\Sigma} = 2 g^{w\wb}R_{w\wb}=2\varkappa =const.
\end{equation}
Note that for genus $g\ne 1$ the Riemann surface $\Sigma$ has area equal to
\begin{equation}\label{2.6}
{\rm{Vol}}(\Sigma) = \int_{\Sigma}\omega_{\Sigma}^{} = \frac{4\pi}{\varkappa}\,(1-g)
\end{equation}
which follows from (\ref{2.5}) and the Gauss-Bonnet theorem.

\smallskip

\noindent
{\bf Four-manifold $M$.} Let us consider a smooth oriented real four-dimensional
manifold $M$ given by a product of two Riemann surfaces of genera $g_1$ and $g_2$,
\begin{equation}\label{2.7}
M=\Sigma_{1}\times \Sigma_{2}\ ,
\end{equation}
with the product metric
\begin{equation}\label{2.8}
\diff s^2_M = \diff s^2_1+\diff s^2_2= 2\,g_{y_1\yb_1}\diff y_1\diff \yb_1 +
2\,g_{y_2\yb_2}\diff y_2\diff \yb_2
\end{equation}
written in local complex coordinates on $\Sigma_{1}$ and $\Sigma_{2}$, respectively.

We consider a principal bundle $P=P(M,$ SO(4)) of orthonormal frames on a Riemannian
4-manifold $M$. For $M=\Sigma_{1}\times \Sigma_{2}$ the holonomy group is reduced
to U(1)$\times$U(1)$\ \subset\ $U(2)$\ \subset\ $SO(4) and for the components
$g^{y_1\yb_1}=1/g_{y_1\yb_1}$ and $g^{y_2\yb_2}=1/g_{y_2\yb_2}$ of inverse metric we have
\begin{equation}\label{2.9}
g^{y_1\yb_1}=e_1^{y_1}\,e_{\1}^{\yb_1}\quad\mbox{and}\quad
g^{y_2\yb_2}=e_2^{y_2}\,e_{\2}^{\yb_2}\ ,
\end{equation}
where $e^{y_1}_1,\ldots , e^{\yb_2}_{\2}$ are unitary (local) frame fields.
We also introduce a basis of type (1,0) and (0,1) vector fields
\begin{equation}\label{2.10}
e_1:=e^{y_1}_1\,\pa_{y_1}\ ,\quad e_2:=e^{y_2}_2\,\pa_{y_2}\ ,\quad
e_{\1}:=e^{\yb_1}_{\1}\,\pa_{\yb_1}\quad\mbox{and}\quad
e_{\2}:=e^{\yb_2}_{\2}\,\pa_{\yb_2}\ ,
\end{equation}
which are sections of the complexified tangent bundle
$T^{\C}M=TM\otimes\C =T^{1,0}M\oplus T^{0,1}M$
with $TM$ associated to the principal bundle $P(M,$ SO(4)).
Dual basis of type (1,0) and (0,1) forms is
\begin{equation}\label{2.11}
\b^1:=e_{y_1}^1\,\diff{y_1}\ ,\quad \b^2:=e_{y_2}^2\,\diff{y_2}\ ,\quad
\b^{\1}:=e_{\yb_1}^{\1}\,\diff{\yb_1}\quad\mbox{and}\quad
\b^{\2}:=e_{\yb_2}^{\2}\,\diff{\yb_2}\ ,
\end{equation}
where $e_{y_1}^1$ etc. are inverse to $e^{y_1}_1$ etc., i.e. $e_{y_1}^1e^{y_1}_1=1$ etc.

Introducing
\begin{equation}\label{2.12}
\r^2_1:=g_{y_1\yb_1}(y_1, \yb_1)\quad\mbox{and}\quad \r^2_2:=g_{y_2\yb_2}(y_2, \yb_2)\ ,
\end{equation}
we obtain
\begin{equation}\label{2.13}
\Gamma^{y_1}_{y_1y_1} = 2\,\pa_{y_1}\log\rho_1\ ,\quad
\Gamma^{y_2}_{y_2y_2} = 2\,\pa_{y_2}\log\rho_2\ ,\quad
\Gamma^{\yb_1}_{\yb_1\yb_1} = 2\,\pa_{\yb_1}\log\rho_1\ ,\quad
\Gamma^{\yb_2}_{\yb_2\yb_2} = 2\,\pa_{\yb_2}\log\rho_2\ ,
\end{equation}
\begin{equation}\label{2.14}
R_{y_1\yb_1} = -\pa_{y_1}\pa_{\yb_1}\log\rho_1^2 =\varkappa_1\, g_{y_1\yb_1}
\quad\mbox{and}\quad
R_{y_2\yb_2} = -\pa_{y_2}\pa_{\yb_2}\log\rho_2^2 =\varkappa_2\, g_{y_2\yb_2}
\end{equation}
with all other components vanishing. For the scalar curvature of $M$ we have
\begin{equation}\label{2.15}
R^{}_M = 2 g^{y_1\yb_1}R_{y_1\yb_1} + 2 g^{y_2\yb_2}R_{y_2\yb_2}= 2(\varkappa_1+\varkappa_2)\ .
\end{equation}

\smallskip

\noindent
{\bf Twistor space of $M$.}  Twistor space of an oriented four-dimensional manifold
$M$ can be defined as the associated bundle
\begin{equation}\label{2.16}
{\Zcal}=P\times_{\rm{SO}(4)}\C P^1
\end{equation}
with the canonical projection
\begin{equation}\label{2.17}
\pi : \Zcal\to M\ .
\end{equation}
Recall that $P$ is the principal SO(4)-bundle of oriented orthonormal frames.
Fibres of the bundle (\ref{2.17}) are two-spheres $S_x^2\cong\C P_x^1\cong\ $SO$(4)/$U(2)
parametrizing complex structures on tangent spaces $T_xM$ at $x\in M$.

Equivalent definition of the twistor bundle (\ref{2.17}) can be obtained by
considering the vector bundle $\Lambda^2 M$ of two-forms on $M$, associated
to the principal bundle $P$. Using projectors on the subspaces of self-dual
$\Lambda^2_+$ and anti-self-dual $\Lambda^2_-$ two-forms, one can split
$\Lambda^2 M$ into the direct sum $\Lambda^2 M = \Lambda^2_+M \oplus \Lambda^2_-M$
of subbundles of self-dual and anti-self-dual two-forms on $M$. Then the twistor
space can be introduced as the unit sphere bundle $\Zcal = S_1(\Lambda^2_-M)$
in the vector bundle $\Lambda^2_-M$. That is why the Levi-Civita connection
on $P$ determines that on $\Lambda^2 M$ and induces a connection on $\Zcal$
 which is the anti-self-dual part $\Gamma_-=(\Gamma_-^i), i=1,2,3$, of the
 Levi-Civita connection~\cite{Atiyah:1978wi}. This connection generates the
 splitting of the tangent bundle $T\Zcal$ into the direct sum $T\Zcal =
 H\oplus V$, where $V={\rm Ker}\,\pi_*$ is the subbundle of vectors tangent to
 fibres $\C P^1$, and $H\cong T\Zcal /V$ consists of vectors horizontal
 with respect to the connection $\Gamma_-$ on $\Zcal$. According to the canonical
 definition, the horizontal lift $\tilde X$ of any vector field $X$ on $M$
 is defined as
\begin{equation}\label{2.18}
\tilde X:= X + (X \lrcorner\, \Gamma_-^i)L_i\ ,
\end{equation}
where $X \lrcorner\, \Gamma_-^i$ denotes the interior product of a vector field
and one-form, and $L_i$'s are vector fields on fibres $\C P^1\hookrightarrow
\Zcal$ which give a realization of the generators of the group SU(2).
By construction, $\tilde X$ is a section of the bundle $H\to\Zcal$, $X$ is a
section of $TM$ and $\pi_*\tilde X =X$.

We lift our frame vector fields (\ref{2.10}) to $\Zcal$, switch to a complex basis
by taking holomorphic parts of vector fields $L_i$ and introduce type (0,1) vector
fields on
$\Zcal$ as
\begin{equation}\label{2.19}
V_{\1}:= \tilde e_{\1} - \la \tilde e_2\ , \ V_{\2}:= \tilde e_{\2} + \la \tilde e_1\quad\mbox{and}\quad
V_{\3}:=\pa_{\bar\la}\ ,
\end{equation}
where in local complex coordinates $y_1, y_2, \la$ on $\Zcal$ ($\la\in\C P^1\hookrightarrow\Zcal$)
we have
\begin{subequations}\label{2.20}
\begin{eqnarray}
\tilde e_1=\rho_1^{-1}\pa_{y_1} - \rho_1^{-1}(\pa_{y_1}\log\rho_1)\la\pa_{\la} \ ,\quad
\tilde e_2=\rho_2^{-1}\pa_{y_2} - \rho_2^{-1}(\pa_{y_2}\log\rho_2)\la\pa_{\la} \ , \\
\tilde e_{\1}= \rho_1^{-1}\pa_{{\yb}_1} + \rho_1^{-1}(\pa_{{\yb}_1}\log\rho_1)\la\pa_{\la}\quad\mbox{and}\quad
\tilde e_{\2}= \rho_2^{-1}\pa_{\yb_2} + \rho_2^{-1}(\pa_{\yb_2}\log\rho_2)\la\pa_{\la} \ .
\end{eqnarray}
\end{subequations}

\noindent
{\bf Almost complex structure on $\Zcal$.} The vector fields (\ref{2.19}) define an {\it almost complex
structure} $\J$ on $\Zcal$ such that
\begin{equation}\label{2.21}
\J(V_{\bar k}) = -\im V_{\bar k}
\end{equation}
for $k=1,2,3$. In fact, the almost complex structure $\J$ defined as above on the twistor space $\Zcal$ of
$M=\Sigma_1{\times}\Sigma_2$ is canonical and does not depend on the choice of local
coordinates~\cite{Atiyah:1978wi}. For commutators of type (0,1) vector fields (\ref{2.19}) we have
\begin{equation}\label{2.22}
[V_{\1}, V_{\2}]=\la\rho_1^{-2}({\pa}_{y_{1}}\rho_1)V_{\1} + \la \rho_2^{-2}({\pa}_{y_{2}}\rho_2)V_{\2} +
2{{\la}^2}({\varkappa_1}+{\varkappa_2})V_{3}\quad\mbox{and}\quad
  {[V_{\1}, V_{\3}]}=0=[V_{\2}, V_{\3}]\ ,
\end{equation}
where $V_3=\pa_\la$ is the (1,0) vector field.

Recall that for integrability of an almost complex structure $\J$ on $\Zcal$ it is necessary and sufficient
that the commutator of any two vector fields of type (0,1) w.r.t. $\J$ is of type (0,1). For
our case  we see from (\ref{2.22}) that $\J$ is integrable - and $\Zcal$ is a complex manifold - if
and only if
\begin{equation}\label{2.23}
\varkappa_1=-\varkappa_2\ ,
\end{equation}
i.e when the scalar curvature (\ref{2.15}) of the K\"ahler manifold $\Sigma_1\times\Sigma_2$ vanishes.
Besides the case $g_1=1=g_2$ (tori) this can be satisfied when one takes $g:=g_1\ge 2$ and $g_2=0$
(two-sphere).

\smallskip

\noindent
{\bf Gravitational instantons.} Recall that the Weyl tensor for the manifold $\Sigma_1\times\Sigma_2$
with equal and opposite scalar curvatures of $\Sigma_1$ and $\Sigma_2$ is self-dual
(or anti-self-dual for the inverse orientation)~\cite{B, LB2}. Such manifolds
$\Sigma_1\times\Sigma_2$ with $\varkappa_1=-\varkappa_2$ are considered as gravitational instantons in
conformal gravity~\cite{Atiyah:1978wi, Gibbons1}.
However, the case $\varkappa_1=\varkappa_2$ is also of interest since such 4-manifolds
$\Sigma_1\times\Sigma_2$ are smooth Einstein spaces which were also considered as gravitational
instantons in Euclidean quantum gravity (see e.g.~\cite{Gibbons2, Gibbons3}).
Note that the twistor space $\Zcal$ of such manifolds is an almost complex manifold with the
nonintegrable almost complex structure $\J$ defined by formulae (\ref{2.19})-(\ref{2.22}).
In what follows we consider the case with
\begin{equation}\label{2.24}
\varkappa_1=\varkappa\in\R\quad\mbox{and}\quad\varkappa_2=\frac{1}{R^2}>0
\end{equation}
including $\varkappa =0$ ($T^2\times S^2$) and the special cases $\varkappa_1=-\varkappa_2$
($\Sigma\times S^2$) and $\varkappa_1=\varkappa_2$ ($S^2\times S^2$) corresponding to both types of
(conformally self-dual and non-self-dual) gravitational instantons.

\vspace{5mm}

\section{Vortices on $\Sigma$ as Yang-Mills instantons on $\Sigma\times\C P^1$}

\noindent
{\bf Riemann sphere.} Let $\Sigma_2=\C P^1\cong S^2$ be the standard two-sphere of constant radius $R$.
In local coordinates on $\C P^1$ the metric reads
\begin{equation}\label{3.1}
\diff s_2^2 = 2\,g_{y\yb}\,\diff y\,\diff \yb=\frac{4R^4}{(R^2+y\yb)^2}\,\diff y\,\diff \yb=
R^2(\diff\th^2+\sin^2\th\,\diff\vp^2)
\end{equation}
for
\begin{equation}\label{3.2}
y:=y_2=R\tan\Bigl(\frac{\th}{2}\Bigr)\exp(-\im\vp )\ ,\quad
\yb:=\yb_2=R\tan\Bigl(\frac{\th}{2}\Bigr)\exp(\im\vp )\ ,\quad 0\le\th < \pi\ ,
\quad0\le\vp\le 2\pi
\end{equation}
Note that (\ref{3.2}) corresponds to the choice of orientation on $S^2$ inverse to the canonical one,
i.e. $y_2=x^3-\im x^4$ and $\yb_2=x^3+\im x^4$ for real local coordinates $x^3$, $x^4$
on $S^2$. That is why for the volume form we have
\begin{equation}\label{3.3}
\omega_2 = R^2\,\sin^2\th\,\diff\th\wedge\diff\vp=
- \frac{2\im\,R^4}{(R^2+y\yb)^2}\,\diff y\wedge\diff \yb=
-\im\,g_{y\yb}\,\diff y\wedge\diff \yb\ ,
\end{equation}
i.e. $\omega_2$ has the inverse sign in comparison with (\ref{2.2}).

For the two-sphere
\begin{equation}\label{3.4}
R_{y\yb}=-\pa_y \pa_{\yb}\log\rho_2^2 = \frac{1}{R^2}\rho^2_2\quad\Rightarrow\quad
\varkappa_2=\frac{1}{R^2}>0\ ,
\end{equation}
and we have
\begin{equation}\label{3.5}
\tilde e_2=\frac{(R^2+y\yb)}{\sqrt{2}\,R^2}\,\pa_y   + \frac{\yb}{\sqrt{2}\,R^2}\,{\la}{\pa}_{\la}
\quad\mbox{and}\quad
\tilde e_{\2}=\frac{(R^2+y{\yb})}{\sqrt{2}\,R^2}\,\pa_{\yb}- \frac{y}{\sqrt{2}\,R^2}\,{\la}{\pa}_{\la} \ .
\end{equation}
For the area of $S^2$ we have
\begin{equation}\label{3.6}
\mbox{Vol}(S^2)=\int_{S^2}\omega_2=4\pi R^2\ ,
\end{equation}
which agrees with the general formula (\ref{2.6}).

\noindent
{\bf Monopole bundles. } Consider the Hermitian line bundle $\Lcal\to\C P^1$
(one-monopole bundle) with a unique SU(2)-equivariant unitary connection
\begin{equation}\label{3.7}
a=-\frac{\im}{2}\,(1-\cos\th )\,\diff\vp = \frac{1}{2(R^2+y\yb)}(\yb\,\diff y - y\,\diff\yb )\ .
\end{equation}
Then on the bundle $\Lcal^n:=(\Lcal)^{\otimes n}$ with $n\in\Z$ we have\footnote{For more
detailed description with transition functions etc. see e.g.~\cite{Popov:2004rt}.}
\begin{equation}
a^{(n)}=n\,a\quad\mbox{and}\quad f^{(n)}:=\diff a^{(n)}=-\frac{nR^2}{(R^2+y\yb)^2}\, \diff y\wedge\diff\yb
=\frac{n}{2\im\,R^2}\,\omega_2\ .
\end{equation}\label{3.8}
The topological charge of this gauge field configuration is given by the first Chern number
(equivalently the degree) of the complex line bundle $\Lcal$,
\begin{equation}\label{3.9}
c_1(\Lcal^n)=\frac{\im}{2\pi}\int_{\C P^1}f^{(n)}=\frac{n}{4\pi R^2}\int_{\C P^1}\omega_2 = n
\end{equation}
This configuration describes $|n|$ Dirac monopoles ($n>0$) or antimonopoles ($n<0$) sitting on the
top of each other.

\smallskip

\noindent
{\bf SU(2)-equivariant gauge potential.} Let $\Ecal\to M$ be a rank-2 SU(2)-equivariant complex
vector bundle over $M=\Sigma\times\C P^1$ with the group SU(2) acting trivially on $\Sigma$ and
in the standard way by SU(2)-isometry on $\C P^1=\ $SU(2)/U(1). Let $\Acal$ be an $su(2)$-valued
equivariant connection on $\Ecal$. The explicit form of such a connection is known
(see e.g.~\cite{Forgacs:1979zs, GarciaPrada:1993qv, Popov:2005ik});
it has the form
\begin{equation}\label{3.10}
\Acal = \begin{pmatrix}\sfrac12A\otimes 1+ 1\otimes a& \sfrac{1}{\sqrt{2}}\phi\otimes\bar\b
\\-\sfrac{1}{\sqrt{2}}\bar\phi\otimes\b&-\sfrac12A\otimes 1 - 1\otimes a\end{pmatrix}
\end{equation}
where $A$ is an Abelian connection on the (Hermitian) complex line bundle $E$ over the
genus $g$ compact Riemann surface $\Sigma$, $a$ is the monopole connection (\ref{3.7}) on the
line bundle $\Lcal\to\C P^1$, $\phi$ is a section of the bundle $E$, $\bar\phi$ is its
complex conjugate and
\begin{equation}\label{3.11}
\b:=\b^2\frac{\sqrt{2}\,R^2\,\diff y}{R^2+y\yb}\quad\mbox{and}\quad
\bar\b:=\b^{\2}=\frac{\sqrt{2}\,R^2\,\diff \yb}{R^2+y\yb}
\end{equation}
are forms on $\C P^1$ of type (1,0) and (0,1) given by (\ref{2.11}).  In local complex coordinates
$z:=y_1=x^1+\im x^2$ and $\zb:=\yb_1=x^1-\im x^2$,  we have $A=A(z,\zb)$ and $\phi =\phi (z, \zb)$.

Note that for any rank$(\Ecal )\ge 2$ the $\C P^1$-dependence is uniquely determined by the
SU(2)-equivariance, by rank of $\Ecal$ and the monopole configuration on
$\C P^1$~\cite{GarciaPrada:1993qv, Popov:2005ik}.
In particular, in the above case of rank$(\Ecal ) =2$ the forms (\ref{3.11}) are the unique
SU(2)-invariant (1,0) and (0,1) forms such that
\begin{equation}\label{3.12}
\diff\bar\b + 2a\wedge\bar\b =0=\diff\b - 2a\wedge\b
\end{equation}
and the K\"ahler (1,1)-form $\omega_2$ on $\C P^1$ is $-\im\b\wedge\bar\b$. The forms $\b$
and $\bar\b$ take values in the bundles $K=\Lcal^2$ and $K^{-1}=\Lcal^{-2}$, respectively.

\smallskip

\noindent
{\bf Field strength tensor.} In local complex coordinates on $\Sigma\times\C P^1$ the calculation
of the curvature $\Fcal$ for $\Acal$ of the form (\ref{3.10}) yields
\begin{equation}\nonumber
\Fcal = \diff\Acal +\Acal\wedge\Acal =\begin{pmatrix}\sfrac12F -
\sfrac12(\sfrac{1}{R^2}-\phi\bar\phi )\b\wedge\bar\b& \sfrac{1}{\sqrt{2}}(\diff\phi + A\phi )
\wedge\bar\b
\\-\sfrac{1}{\sqrt{2}}(\diff\bar\phi - A\bar\phi )\wedge\b&-\sfrac12F +
\sfrac12(\sfrac{1}{R^2}-\phi\bar\phi )\b\wedge\bar\b\end{pmatrix}=
\end{equation}
\begin{equation}\label{3.13}
=\Fcal_{z\bar z}\,\diff z\wedge\diff\bar z + \Fcal_{zy}\,\diff z\wedge\diff y +
\Fcal_{z\bar y}\,\diff z\wedge\diff\bar y + \Fcal_{\bar z y}\,\diff \bar z\wedge\diff y
+\Fcal_{\bar z\bar y}\,\diff \bar z\wedge\diff\bar y +\Fcal_{y\bar y}\,\diff y\wedge\diff\bar y
\end{equation}
with the non-vanishing field strength components
\begin{equation}\label{3.14}
\Fcal_{z\zb}=\frac{1}{2} F_{z\zb}\,\s_3\ ,\quad
\Fcal_{y\yb}=-\frac{\r_2^2}{2} (\frac{1}{R^2}-\phi\bar\phi )\,\s_3\ ,
\end{equation}
\begin{equation}\label{3.15}
\Fcal_{\zb\yb}=\frac{\rho_2}{\sqrt{2}}\,(\pa_{\zb}\phi + A_{\zb}\phi )\,\s_+
\ ,\quad
\Fcal_{z\yb}=\frac{\rho_2}{\sqrt{2}}\,(\pa_{z}\phi + A_{z}\phi )\,\s_+\ ,
\end{equation}
\begin{equation}\label{3.16}
\Fcal_{zy}=-\frac{\rho_2}{\sqrt{2}}\,(\pa_{z}\bar\phi - A_{z}\bar\phi )\,\s_-
\ ,\quad
\Fcal_{\zb y}=-\frac{\rho_2}{\sqrt{2}}\,(\pa_{\zb}\bar{\phi} - A_{\zb}\bar{\phi} )\,\s_-
\ ,
\end{equation}
where
\begin{equation}\label{3.17}
\s_3=\begin{pmatrix}1&0\\0&-1\end{pmatrix}\ ,\quad \s_+=\begin{pmatrix}0&1\\0&0\end{pmatrix}
\quad\mbox{and}\quad\s_-=\begin{pmatrix}0&0\\1&0\end{pmatrix}\ .
\end{equation}
In (\ref{3.14}) we have defined $F=\diff A=F_{z\zb}\,\diff z\wedge \diff\zb=(\pa_z A_{\zb}
-\pa_{\zb}A_z)\,\diff z\wedge \diff\zb$ for $A=A_z\diff z+A_{\zb}\diff\zb$.

\smallskip

\noindent
{\bf Reduction of the Yang-Mills functional.} The dimensional reduction of the Euclidean
Yang-Mills equations from $\Sigma\times\C P^1$ to $\Sigma$ can be seen at the level of the
Yang-Mills Lagrangian. Substituting (\ref{3.14})-(\ref{3.16}) into the standard Yang-Mills
functional and performing  the integral over $\C P^1$, we arrive at the action
\begin{eqnarray}\nonumber
S&=&-\frac{1}{8\pi^2}\int_M\tr\, (\Fcal\wedge *\Fcal )=
-\frac{1}{8\pi^2}\int_{\Sigma\times\C P^1}\diff^4 x\, \sqrt{\det (g_{\r\s})}\, \tr\,
(\Fcal_{\mu\nu}\Fcal^{\mu\nu})=\\
&=&\frac{R^2}{2\pi}\int_{\Sigma}\omega_\Sigma^{}\,\Bigl\{(g^{z\zb})^2\,(F_{z\zb})^2 +
2g^{z\zb}(D_z\phi\,D_{\zb}\bar\phi + D_{\zb}\phi\,D_{z}\bar\phi )
+ (\frac{1}{R^2}-\phi\bar\phi )^2\Bigr\}\ ,\label{3.18}
\end{eqnarray}
where
\begin{equation}\label{3.19}
\omega_\Sigma^{}\equiv\omega_1=\im\, g_{z\zb}\,\diff z\wedge\diff\zb
\end{equation}
and $x^\m$ are real local coordinates\footnote{One can take e.g. $y_1=z=x^1+\im x^2$
and $y_2=y=x^3-\im x^4$} on $M=\Sigma\times\C P^1, \m ,\n ,...=1,...,4$.

For the functional (\ref{3.18}), using the standard Bogomolny arguments~\cite{Bogomolny:1975de},
one can obtain the inequality
\begin{equation}\label{3.20}
S=\frac{R^2}{2\pi}\,\int_{\Sigma}\im\,
\diff z\wedge\diff\zb\,\Bigl\{g^{z\zb}\,\bigl(F_{z\zb} +
g_{z\zb}\,(\phi\bar\phi -\frac{1}{R^2})\bigr)^2
+2D_{\zb}\phi\,\overline{(D_{\zb}\phi )}\,\Bigr\}+
\frac{\im}{2\pi}\,\int_{\Sigma}F\ \ge\ N\ ,
\end{equation}
where
\begin{equation}\label{3.21}
N=-c_2(\Ecal)=-\frac{1}{8\pi^2}\,\int_{\Sigma\times\C P^1}\tr\, (\Fcal\wedge\Fcal )=
\frac{\im}{2\pi}\,\int_\Sigma\, F=c_1(E)
\end{equation}
is the vortex number. In the derivation of (\ref{3.20}) it is assumed that $N\ge 0$
and similar inequality can be obtained for $N\le 0$. Thus, the (minus) second Chern
number of the SU(2)-equivariant connection $\Acal$ on the rank-2 bundle $\Ecal$
over $\Sigma\times\C P^1$ equals to the first Chern number of the connection $A$
on the line bundle $E$ over $\Sigma$.

\smallskip

\noindent
{\bf Vortex equations on $\Sigma$.} If $N\ge 0$, then (\ref{3.20}) is an equality if and
only if the Yang-Mills field $\Fcal$ on $\Sigma\times\C P^1$ is self-dual,
\begin{equation}\label{3.22}
*\Fcal=\Fcal\ ,
\end{equation}
and substitution of (\ref{3.14})-(\ref{3.16}) shows that Eqs. (\ref{3.22}) are equivalent
to the vortex equations on $\Sigma$,
\begin{subequations}\label{3.23}
\begin{eqnarray}
F_{z\zb}=g_{z\zb}\,(\frac{1}{R^2}-\phi\bar\phi )\quad&\Leftrightarrow&\quad
\im\,F=(\frac{1}{R^2}-\phi\bar\phi )\,\omega_\Sigma^{}\ ,\\
\pa_{\zb}\phi + A_{\zb}\phi =0\quad&\Leftrightarrow&\quad\bar\pa_A\phi =0\ .
\end{eqnarray}
\end{subequations}
Note that the case $N\le 0$ corresponds to the anti-self-dual Yang-Mills equations
$*\Fcal=-\Fcal$ which reduce to the anti-vortex equations
\begin{equation}
F_{z\zb}=-g_{z\zb}\,(\frac{1}{R^2}-\phi\bar\phi )\quad\mbox{and}\quad
\pa_{z}\phi + A_{z}\phi =0\ .
\end{equation}
For this reduction one should simply change $y\to\yb$ in (\ref{3.10}) and (\ref{3.11})
which is equivalent to the change of the orientation $x^4\to -x^4$ of $M$. It is well
known that self-duality is transformed into anti-self-duality under the change of
orientation on $M$.

{}From (\ref{3.21}) we see that the instanton number of the SU(2)-equivariant Yang-Mills field on
$\Sigma\times\C P^1$ is equal to the number $N\ge 0$ of vortices on the Riemann surface
$\Sigma$. Furthermore, for $g\ne 1$ from (\ref{3.23}) it follows that
\begin{equation}\label{3.24}
\frac{\im}{2\pi}\,\int_{\Sigma}F + \int_{\Sigma}\phi\bar\phi\,\omega_\Sigma^{}=
\frac{1}{2\pi R^2}\,\int_{\Sigma}\omega_\Sigma^{}=\frac{2}{\varkappa R^2}\, (1-g)
\end{equation}
and we obtain (cf.~\cite{Bradlow:1990ir}) the inequality
\begin{equation}\label{3.25}
N \le\frac{2}{\varkappa R^2 }\, (1-g)\ .
\end{equation}
For $g=1$ we have $N\le\ $Vol$(T^2)/2\pi R^2$. Recall that in our derivation of the vortex
equations (\ref{3.23}) the parameter $\varkappa=\sfrac12\,R^{}_\Sigma$ is not fixed.
In particular, one can take $g=0$ and $g=1$ and obtains vortices on the sphere $S^2$ or
torus $T^2$.

\vspace{5mm}

\section{Twistor description of vortex equations}

In Section 3 we discussed the relation between vortices on Riemann surfaces $\Sigma$ and
Yang-Mills instantons on the manifolds $\Sigma\times\C P^1$. Here, we push this correspondence
further and show that to any solution $(A, \phi )$ of vortex equations on $\Sigma$ there
corresponds a connection $\hat\Acal$ defining a pseudo-holomorphic structure on a rank-2
complex vector bundle $\hat\Ecal$ over an almost complex twistor space $\Zcal$ for
$\Sigma\times\C P^1$. We also discuss the integrability of this pseudo-holomorphic structure
and the corresponding vortex equations.

\smallskip

\noindent
{\bf Pull-back to the twistor space.} Consider an SU(2)-equivariant rank-2 complex vector bundle
$\Ecal\to\Sigma\times\C P^1$ with a connection $\Acal$ described in Section 3. Using the projection
(2.17), we pull $\Ecal$ back to a bundle $\hat\Ecal :=\pi^*\Ecal$ over $\Zcal$:
\begin{equation}\label{4.1}
\begin{CD}
\hat\Ecal@>>>\Zcal\\
@.@VV\pi V\\
\Ecal@>>>M
\end{CD}
\end{equation}
In accordance with the definition of the pull-back, the connection $\hat\Acal :=\pi^*\Acal$
on $\hat\Ecal$ is flat along the fibres $\C P^1_x$ of the bundle $\pi : \Zcal\to M$ and we can set
the components $\hat\Acal_\la$ and $\hat\Acal_{\bar\la}$ of the restriction of $\hat\Acal$ to
$\C P^1_x\hookrightarrow\Zcal$ equal to zero. Thus, restrictions of smooth vector bundle $\hat\Ecal$
to fibres $\C P^1_x$ of the projection $\pi$ are naturally holomorphic and holomorpically trivial
for each $x\in M=\Sigma\times\C P^1$.

Note that a lift of the generators of the group SU(2) acting on $\C P^1=\ $SU(2)$/$U(1) to $\Zcal$ can
be defined analogously to the flat Euclidean case~\cite{Legare}. Namely, we lift the SU(2)-generators to
$\Zcal$ so that an almost complex structure on $\Zcal$ is invariant under the action of
the group SU(2) generated by these lifted vector fields. Then one can impose SU(2)-equivariance
conditions on the vector bundle $\hat\Ecal\to \Zcal$ and connection $\hat\Acal$ on $\hat\Ecal$.

\smallskip

\noindent
{\bf Connection along $T^{0,1}\Zcal$.} Recall that the twistor space $\Zcal$ of $M$ has a canonical
almost complex structure $\J$~\cite{Atiyah:1978wi}. In the considered case it is defined by the type (0,1)
vector fields (\ref{2.19}). Hence, we can introduce a (0,1) part $\hat\nabla^{0,1}$ of the covariant
derivative $\hat\nabla$ on $\hat\Ecal$ by formulae
\begin{subequations}\label{4.2}
\begin{eqnarray}
\hat\nabla_{V_{\1}}^{}&\equiv&V_{\1}+\hat\Acal_{V_{\1}}:= \tilde e_{\1}-\la\tilde e_{2}+\Acal_{\1}-\la\Acal_2\ ,\\
\hat\nabla_{V_{\2}}^{}&\equiv&V_{\2}+\hat\Acal_{V_{\2}}:=\tilde e_{\2}+\la\tilde e_{1}+\Acal_{\2}+\la\Acal_1\ ,
\end{eqnarray}
\end{subequations}
\begin{equation}\label{4.3}
\hat\nabla_{V_{\3}}^{}\equiv V_{\3}+\hat{\Acal}_{V_{\3}}:=\pa_{\bar\la}\ ,
\end{equation}
where $\hat\nabla_X$ denotes the covariant derivative along the vector field $X$.
The components $\Acal_1$ etc. can be extracted from (\ref{3.10}),
\begin{equation}\label{4.4}
{\Acal}_1=\sfrac12\,A_1\,\sigma_3\ ,\quad \Acal_{\1}=\sfrac12\, A_{\1}\,\sigma_3\ ,
\quad {\Acal}_2=a_2\,\sigma_3 - \frac{\bar\phi}{\sqrt{2}}\,\sigma_-\quad\mbox{and}\quad
{\Acal}_{\2}=a_{\2}\,\sigma_3 + \frac{\phi}{\sqrt{2}}\,\sigma_+\ ,
\end{equation}
where $\Acal_1=e^z_1\Acal_z$, $\Acal_{\1}=e^{\zb}_{\1}\Acal_{\zb}$, $\Acal_2=e^y_2\Acal_y$,
$\Acal_{\2}=e^{\yb}_{\2}\Acal_{\yb}$ and similarly for $A_1$, $A_{\1}$ and $a_2$, $a_{\2}$.

\smallskip

\noindent
{\bf Pseudo-holomorphic bundles.}  Let us consider an almost complex manifold $(Q,\J)$
and a complex vector bundle $\cal V$ over $Q$ endowed with a connection $\hat\Acal$.
According to Bryant~\cite{Bryant}, a connection $\hat\Acal$ on $\cal V$  is said to define
a pseudo-holomorphic structure on $\cal V$ if it has curvature $\hat\Fcal$ of type
(1,1). In principle, one can also endow $\cal V$ with a Hermitian metric and
choose $\hat\Acal$ to be compatible with the Hermitian  structure on $\cal V$~\cite{Bryant}.
If, in addition, $\eta$ is an almost Hermitian metric on $(Q,\J)$ and
\begin{equation}\label{4.4'}
\tr_\eta (\hat\Fcal )= \im \g\,\mbox{Id}_{\hat\Ecal}\quad\mbox{with}\quad\g\in\R\ ,
\end{equation}
 the connection $\hat\Acal$  is said to be pseudo-Hermitian-Yang-Mills. Here, we will
consider the equations\footnote{Equations (\ref{4.4'}) together with $\hat\Fcal^{0,2}=0$
can also be called pseudo-Donaldson-Uhlenbeck-Yau equations (cf.~\cite{DUY}) with the prefix
``pseudo-", since $Q$ is not a K\"ahler and not even a complex manifold.}
(\ref{4.4'}) with $\g =0$.

Consider the SU(2)-equivariant complex vector bundle $\hat\Ecal = \pi^*\Ecal$ from
(\ref{4.1}) with a connection $\hat\Acal = \pi^*\Acal$. In~\cite{Atiyah:1978wi}, it was shown that
pulling-back a real structure $\tau : \Zcal\to\Zcal$ to $\hat\Ecal$, one can endow
$\hat\Ecal$ with a Hermitian structure. Then pseudo-holomorphicity of the bundle
$\hat\Ecal$ is equivalent to the equations
\begin{equation}\label{4.5}
\hat\Fcal^{0,2}=0\quad\Leftrightarrow\quad
\hat\Fcal (V_{\bar\imath}, V_{\bar\jmath})=[\hat\nabla_{V_{\bar\imath}}, \hat\nabla_{V_{\bar\jmath}}]-
\hat\nabla_{[V_{\bar\imath}, V_{\bar\jmath}]}=0\ ,
\end{equation}
where $\hat\nabla_{V_{\bar\imath}}$ for $i=1,2,3$ are given in (\ref{4.2}) and (\ref{4.3}).

\smallskip

{\bf Pseudo-holomorphicity and instantons.}  After simple calculations we see that the
only nontrivial components of the tensor $\hat\Fcal^{0,2}$ reads
\begin{equation}\label{4.6}
\hat\Fcal(V_{\1}, V_{\2}) =
\Fcal_{\1\2}-\la\, (\Fcal_{1\1}+\Fcal_{2\2}) +\la^2\,\Fcal_{12}\ ,
\end{equation}
where
\begin{subequations}\label{4.7}
\begin{eqnarray}
\Fcal_{\1\2}&=&e_{\1}\Acal_{\2}- e_{\2}\Acal_{\1} + [\Acal_{\1},\Acal_{\2}]=e_{\1}^{\zb}e_{\2}^{\yb}
\Fcal_{\zb\yb}\ ,\\
\Fcal_{12}&=&-(\Fcal_{\1\2})^{\+}=e_{1}\Acal_{2}- e_{2}\Acal_{1} + [\Acal_{1},\Acal_{2}]=e_{1}^{z}e_{2}^{y}
\Fcal_{zy}\ ,\\
\Fcal_{1\1}&=&e_{1}\Acal_{\1}- e_{\1}\Acal_{1} + [\Acal_{1},\Acal_{\1}]-
\r^{-1}_1(e_{\1}\r_1)\Acal_1 +  \r^{-1}_1(e_{1}\r_1)\Acal_{\1}=g^{z\zb}\Fcal_{z\zb}\ ,\\
\Fcal_{2\2}&=&e_{2}\Acal_{\2}- e_{\2}\Acal_{2} + [\Acal_{2},\Acal_{\2}]-
\r^{-1}_2(e_{\2}\r_2)\Acal_2 + \r^{-1}_2(e_{2}\r_2)\Acal_{\2}=g^{y\yb}\Fcal_{y\yb}\ .
\end{eqnarray}
\end{subequations}
Thus, we see that equations (\ref{4.5}) on $\Zcal$, defining pseudo-holomorphic structure on
$\hat\Ecal$, reduce to the equation
\begin{equation}\label{4.8}
\hat\Fcal(V_{\1}, V_{\2}) =
\Fcal_{\1\2}-\la\,(\Fcal_{1\1}+\Fcal_{2\2}) +\la^2\,\Fcal_{12}=0\ ,
\end{equation}
which is equivalent to the SDYM equations on $\Sigma\times\C P^1$,
\begin{equation}\label{4.9}
\Fcal_{\zb\yb}=0\ ,\quad
g^{z\zb}\Fcal_{z\zb} + g^{y\yb}\Fcal_{y\yb}=0
\quad\mbox{and}\quad
\Fcal_{zy}=0\ .
\end{equation}
Conversely, every pseudo-holomorphic SU(2)-equivariant vector bundle
$\hat\Ecal$ over $\Zcal$ such that it is holomorphically trivial
on each $\C P^1_x\hookrightarrow\Zcal$, $x\in\Sigma\times\C P^1$, is the
pull-back to $\Zcal$ of an SU(2)-equivariant bundle $\Ecal$ with a self-dual
connection on $\Sigma\times\C P^1$.\footnote{This can be straightforwardly
generalized to a correspondence between Hermitian vector bundles $\Ecal$
with self-dual connections on an arbitrary oriented 4-manifold $M$ and
pseudo-holomorphic bundles $\hat\Ecal$ over an almost complex twistor space
$\Zcal$ of $M$. This generalization of the Theorem 5.2 in~\cite{Atiyah:1978wi}
will be considered elsewhere.}
Recall that equations (\ref{4.9}) are equivalent in turn to vortex equations on
$\Sigma$ as discussed in Section 3.

\smallskip

\noindent
{\bf Pseudo-Hermitian-Yang-Mills equations on $\Zcal$.} One can check by direct
calculation that the connection $\hat\Acal = \pi^*\Acal$ satisfies the equations
\begin{equation}\label{4.10}
\hat\Fcal(V_{1}, V_{\1}) +
\hat\Fcal(V_{2}, V_{\2})+ \hat\Fcal(V_{3}, V_{\3})=0\ ,
\end{equation}
which also reduce to the SDYM equations (\ref{4.9}). Together, (\ref{4.8}) and
(\ref{4.10}) constitute pseudo-Hermitian-Yang-Mills equations on $\Zcal$.
However, in our concrete case of twistor space $\Zcal$
equation (\ref{4.10}) does not impose additional restrictions on $\hat\Acal$
in comparison with (\ref{4.8}). This is in conformity with~\cite{Bryant}.

Summarizing our above discussion, we have established the diagram
\begin{eqnarray}
&\begin{matrix}\mbox{pseudo-Hermitian-Yang-Mills}\\\mbox{equations on the twistor}\\
\mbox{space $\Zcal$ of $\Sigma\times\C P^1$}\end{matrix}&\nonumber\\
 \swarrow\!\nearrow&&\nwarrow\!\searrow\label{4.11}\\
\mbox{vortex equations on $\Sigma$}&\longleftrightarrow&
\begin{matrix}\mbox{Yang-Mills instanton}\\
\mbox{equations on $\Sigma\times\C P^1$}\end{matrix}\nonumber
\end{eqnarray}
describing equivalent theories defined on different spaces. Furthermore, there are bijections
between the moduli spaces of solutions to all three types of equations mentioned in (\ref{4.11}).

\smallskip

\noindent
{\bf Integrable case.} In the general case of the twistor space $\Zcal$ of $M=\Sigma\times\C P^1$,
 the subbundle $T^{0,1}\Zcal$ of $T^{\C}\Zcal$ is not integrable since the distribution of vector
fields of type (0,1) is not closed under the Lie bracket. However, for $\Sigma$ with genus $g\ge 2$,
one can always rescale the metric on $\Sigma$ to fulfill the condition
\begin{equation}\label{4.12}
\varkappa =-\frac{1}{R^2}\quad\Rightarrow\quad R_M=R_{\Sigma}+R_{S^2}^{}=0\ .
\end{equation}
In this case the almost complex structure $\J$ on the twistor space $\Zcal$ of $M$ becomes integrable
-distribution of (0,1) vector fields (\ref{2.19}) is closed under the Lie bracket - and the
pseudo-holomorphic bundle $\hat\Ecal\to\Zcal$ becomes holomorphic. In other words, the almost complex
structure on $\hat\Ecal$ defined by (0,1)-type connection  (\ref{4.2}) and (\ref{4.3}) becomes
integrable and one can introduce holomorphic sections of $\hat\Ecal$.

\smallskip

\noindent
{\bf Linear system.} Recall that a (local) section $\chi$ of the complex vector bundle $\hat\Ecal$
is said to be holomorphic if
\begin{equation}\label{4.13}
\hat\nabla^{0,1}\chi =0\quad\Leftrightarrow\quad
\hat\nabla_{V_{\1}}\chi =\hat\nabla_{V_{\2}}\chi =\hat\nabla_{V_{\3}}\chi =0\ .
\end{equation}
Accordingly, the bundle $\hat\Ecal\to\Zcal$ is said to be holomorphic if
equations (\ref{4.13}) are compatible, i.e. the (0,2) components of the curvature
$\hat\Fcal$ of the bundle $\hat\Ecal$ vanish. This condition yields equations
(\ref{4.5}) (and (\ref{4.9})) but now with vector fields $V_{\bar\imath}$ which are closed
under the Lie bracket.

Let us now introduce a $2\times 2$ matrix $\psi$ of fundamental solutions of eqs. (\ref{4.13}),
i.e. such that columns of $\psi$ form smooth frame fields for $\hat\Ecal$. Then we obtain
two linear equations (Lax pair)
\begin{subequations}\label{4.14}
\begin{eqnarray}
\hat\nabla_{V_{\1}}\psi (x, \la )\equiv (V_{\1} + \hat\Acal_{V_{\1}})\psi =0\ ,\\
\hat\nabla_{V_{\2}}\psi (x, \la )\equiv (V_{\2} + \hat\Acal_{V_{\2}})\psi =0\ ,
\end{eqnarray}
\end{subequations}
since the third linear equation $\hat\nabla_{V_{\3}}\psi =0$ is trivially solved for $\psi$
which does not depend on $\bar\la$. Here $\la\in U_+=\C P^1\backslash\{\infty\}\subset\C P^1$.

\smallskip

\noindent
{\bf Explicit Lax pair.} In our case, the explicit form of the Lax pair (\ref{4.14})
reads
\begin{subequations}\label{4.15}
\begin{eqnarray}
\left[\tilde{e}_{\1}+{\frac{1}{2}}\, A_{\1}\,\s_3-{\la}\, (\tilde{e}_2 + a_2\,{\s}_3 -{\frac{\bar{\phi}}{{\sqrt{2}}}}\,{\s}_-)\right]\,
\psi =0 \ ,\\
\left[\tilde e_1 + \frac{1}{2} A_1\,\s_3 + \frac{1}{\la}\,(\tilde{e}_{\2} + a_{\2}\,\s_3 + {\frac{{\phi}}{{\sqrt{2}}}}\,\sigma_{+})\right]\,\psi =0\ ,
\end{eqnarray}
\end{subequations}
where $\tilde e_1, \tilde e_2,\tilde e_{\1}$ and $\tilde e_{\2}$ are written down in (\ref{2.20}) and (\ref{3.5}),
and $a_2, a_{\2}$ are given in (\ref{3.7}). By direct calculation, one can see that the compatibility conditions
of the linear equations (\ref{4.15}),
\begin{equation}\label{4.16}
[\hat\nabla_{V_{\1}},\hat\nabla_{V_{\2}}]\,\psi =0\ ,
\end{equation}
are equivalent to the vortex equations (\ref{3.23}) on $\Sigma$. Note that for span$\{V_{\bar\imath}\}$
closed under the Lie bracket, (\ref{4.16}) is equivalent  to (\ref{4.5}) but this is not true for the
nonintegrable distribution $T^{0,1}\Zcal$.

\smallskip

\noindent
{\bf Riemann-Hilbert problems.} Let us consider an open subset $\cal U$ of the manifold $\Sigma\times S^2$,
restrict $\Zcal\mid_{\cal U}\cong{\cal U}\times \C P^1$ and consider a two-set open covering $\{U_+,U_-\}$
of the fibre $\C P^1$ in $\pi :{\cal U}\times \C P^1\to{\cal U}, U_+=\C P^1\backslash\{\infty\}$ and
$U_-=\C P^1\backslash\{0\}$. Then the restriction of $\hat\Ecal$ to $\Zcal\mid_{\cal U}$ will be described by
a transition $2\times 2$ matrix $f_{+-}$ on ${\cal U}\times U_+\cap U_-$, holomorphic in $\la$, whose
restriction to $U_+\cap U_-\hookrightarrow \Zcal\mid_{\cal U}$ is splitted,
\begin{equation}\label{4.17}
f_{+-}=\psi_+^{-1}(x,\la )\psi_-(x,\la )\ ,
\end{equation}
into smooth $2\times 2$ matrices $\psi_+$ and $\psi_-$ which are holomorphic in $\la\in U_+$ and $\la\in U_-$,
respectively. The splitting (\ref{4.17}) can be considered as a solution of a parametric Riemann-Hilbert problem
and the matrix-valued function $\psi$ in (\ref{4.14})-(\ref{4.16}) can be identified with $\psi_+(x,\la )$.
Thus, one can apply various well-known methods (twistor approach, dressing method etc.) to solving
vortex equations on $\Sigma$ with the help of the Lax pair (\ref{4.15}).

\smallskip

\noindent
{\bf Existence of solutions.} Note that, in general, there is a topological obstruction to the existence
of $N$-vortex solution on a compact Riemann surface $\Sigma$~\cite{Bradlow:1990ir}. In particular,
for $g>1$, from
(\ref{3.25}) and (\ref{4.12}) we obtain that solutions of the vortex equations (\ref{3.23}) exist if
\begin{equation}\label{4.18}
N\le 2(g-1)
\end{equation}
since (\ref{4.12}) fixes the area of $\Sigma$ in terms of $g$ and $R^2$. Thus, for
\begin{equation}\label{4.19}
N > 2(g-1)
\end{equation}
the vortex equations on $\Sigma$ written  in the form (\ref{4.16}) will have no solutions.
However, one can rescale the metric on $\Sigma$, $g_{z\zb}\to t^2g_{z\zb}$, and obtain inequalities
\begin{equation}\label{4.20}
R^t_{\Sigma}=-\frac{2}{t^2R^2}<-R_{S^2}=-\frac{2}{R^2}\quad\mbox{for}\quad t^2>1\ ,
\end{equation}
\begin{equation}\label{4.21}
N\le 2t^2(g-1)\ .
\end{equation}
For any $N$ the condition (\ref{4.21}) can be satisfied for sufficiently large $t$ and then the
moduli space of vortices will be nonempty. However, this rescaling of metric on $\Sigma$ leading to
(\ref{4.20}) brings us back to the case of pseudo-holomorphic bundles $\hat\Ecal$ over $\Zcal$, i.e.
to the case of nonintegrable almost complex structures. This is the price to be paid for having
nonempty vortex moduli space when $N>2(g-1)$. The value $N=2(g-1)$ of the vortex
number separates the holomorphic and nonholomorphic cases.

\vspace{5mm}

\section{Conclusions}

In this paper, we have shown the equivalence between vortex equations on a compact Riemann surface
$\Sigma$, the Yang-Mills instanton equations on $\Sigma\times\C P^1$ and pseudo-Hermitian-Yang-Mills
equations on the twistor space $\Zcal$ of $\Sigma\times\C P^1$, summarized in the diagram (\ref{4.11}).
We have shown that in a special case, when the twistor geometry becomes integrable (holomorphic), the
vortex equations on $\Sigma$ appear as the commutator (\ref{4.16}) of two auxiliary linear differential
operators (with a `spectral' parameter) having clear geometric sense. This brings us to a situation when
one can, in principle, apply methods of integrable systems to finding solutions of vortex equations.

We considered vortices in the Abelian Higgs model. There are various non-Abelian generalizations of
this model (see e.g.~\cite{Bradlow1}-\cite{group1} and references therein). Results of this paper can
be extended to the non-Abelian case.
The simplest generalization can be obtained if one replaces  the  scalar function $\phi$ in (\ref{3.10})
by $p\times q$ matrix and substitutes ${1}_p\otimes a$ and $-{1}_q\otimes a$ for $1\otimes a$ and
$-1\otimes a$, respectively. We will consider this case in more detail elsewhere.

The constructions of this paper can also be generalized to $\Ncal$-extended supersymmetric SDYM
theory~\cite{group2, group3} defined on $\Sigma\times\C P^1$ together with a reduction to supersymmetric
vortex equations on $\Sigma$. In this case vortices on $\Sigma$ will correspond to supersymmetric
Yang-Mills instantons in the background of gravitational instantons $\Sigma\times\C P^1$ of conformal
(super)gravity. It is of interest since conformal supergravity interacting with Yang-Mills supermultiplets
 arises in the twistor string theory proposed recently~\cite{Witten:2003nn, Berkovits:2004jj}.
Vortices on $\Sigma$ represent the simplest and very natural interaction of Yang-Mills instantons
and gravitons in conformal gravity which cannot be reduced to Einstein gravity.
These Yang-Mills/gravity configurations could be a good test background for calculation of open/closed
twistor string amplitudes especially because the moduli spaces of Riemann surfaces and vortices are studied
very well. It would be also interesting to look at our vortices from the cosmic string perspective.

\smallskip

\section*{Acknowledgements}

\noindent
This work was supported in part by the Deutsche Forschungsgemeinschaft and the Russian
Foundation for Basic Research (grant 08-01-00014-a).

\smallskip
%\newpage

\end{document}